\documentstyle[12pt]{article}

\def\etal{{\em et al}}
\def\QG{{\rm QG}}
\def\P{{\rm P}}
\def\lappeq{\mathrel{\rlap{\raise.5ex\hbox{$<$}}{\lower.5ex\hbox{$\sim$}}}}

\begin{document}

\begin{flushright}
ACT-18/97 \\
CTP-TAMU-49/97 \\
OUTP-97-73-P \\
NEIP-97-013\\
astro-ph/9712103 \\
$~$ \\
accepted for publication in {\it Nature}
\end{flushright}

\begin{centering}
\bigskip
{\Large \bf Potential Sensitivity of Gamma-Ray Burster Observations 
             to Wave Dispersion in Vacuo} \\
\bigskip
{\bf G. Amelino-Camelia$^{a,b}$},
{\bf John Ellis$^{c}$},
{\bf N.E. Mavromatos$^{a}$}, \\
{\bf D.V. Nanopoulos}$^{d}$ and 
{\bf Subir Sarkar}$^{a}$ \\
\bigskip \noindent 
$^a$ Theoretical Physics, University of Oxford, 1 Keble Road, Oxford
     OX1 3NP, UK \\
$^b$ Institut de Physique, Universit\'e de Neuch\^atel,
     CH-2000 Neuch\^atel, Switzerland \\
$^c$ Theory Division, CERN, CH-1211, Geneva, Switzerland, \\
$^d$ Academy of Athens, Chair of Theoretical Physics, Division of Natural
     Sciences, 28~Panepistimiou Ave., Athens GR-10679, Greece,
     Center for Theoretical Physics, Dept. of Physics, Texas A \& M
     University, College Station, TX 77843-4242, USA and Astroparticle
     Physics Group, Houston Advanced Research Center (HARC), The Mitchell
     Campus, Woodlands, TX 77381, USA \\
\end{centering}
\vspace{1cm}

\baselineskip = 24pt

\noindent
{\bf The recent confirmation that at least some gamma-ray bursters
(GRBs) are indeed at cosmological
distances~\cite{paradis,groot,metzeger1,metzeger2} raises the
possibility that observations of these could provide interesting
constraints on the fundamental laws of physics. Here we demonstrate
that the fine-scale time structure and hard spectra of GRB emissions
are very sensitive to the possible dispersion of electromagnetic waves
{\it in vacuo} with velocity differences $\delta\,v\sim\,E/E_{\QG}$,
as suggested in some approaches to quantum gravity. A simple estimate
shows that GRB measurements might be sensitive to a dispersion scale
$E_{\QG}$ comparable to the Planck energy scale
$E_{\P}\sim\,10^{19}$~GeV, sufficient to test some of these theories,
and we outline aspects of an observational programme that could
address this goal.}

\medskip \par 
It has been suspected that at least some gamma-ray bursters (GRBs) are
at cosmological distances, on the basis of their isotropy and the
deviation of their brightness distribution from the Euclidean form at
the faint end~\cite{review}. The first direct evidence for this was
provided by the discovery~\cite{paradis} of an extended faint optical
source, probably a galaxy, coincident with GRB~970228. Subsequently,
the detection of both Mg~II absorption lines and O~II emission lines
in the optical afterglow of GRB~970508 has allowed a measurement of
its distance, thanks to a precise determination of its redshift:
$z=0.835\pm0.001$~\cite{metzeger1,metzeger2}. Such a cosmological
distance, combined with the short time structure seen in emissions
from some GRBs~\cite{review}, bestows on GRBs unique features as
`laboratories' for fundamental physics as well as for astrophysics.

\medskip \par
Our interest is in the search for possible {\em in vacuo} dispersion,
$\delta\,v\sim\,E/E_{\QG}$, of electromagnetic radiation from GRBs,
which could be sensitive to a type of candidate quantum-gravity effect
that has been recently considered in the particle-physics literature.
This candidate quantum-gravity effect would be induced 
by a deformed dispersion relation for photons of the 
form $c^2{\bf p}^2 = E^2 \,[1+f(E/E_{\QG})]$, 
where $E_{\QG}$ is an effective quantum-gravity
energy scale and $f$ is a model-dependent function of the
dimensionless ratio $E/E_{\QG}$.  In quantum-gravity scenarios in
which the Hamiltonian equation of motion ${\dot
x}_i\,=\partial\,H/\partial\,p_i$ is still valid at least
approximately, as in the frameworks discussed later, such a deformed
dispersion relation would lead to energy-dependent velocities
$c+\delta\,v$ for massless particles, with implications for all the
electromagnetic signals that we receive from astrophysical objects at
large distances. At small energies $E \ll\,E_{\QG}$, we expect that
a series expansion of the dispersion relation should be applicable:
$c^2{\bf p}^2 = E^2 \left[1 + \xi {E /E_{\QG}}+ {\cal
O}({E^2/E_{\QG}^2})\right]$, where $\xi = \pm 1$ is a sign ambiguity
that would be fixed in a given dynamical framework. Such a series
expansion would correspond to energy-dependent velocities
\begin{eqnarray}
 v = {\partial E \over \partial p} \sim
 c  \left( 1 - \xi {E \over E_{\QG}} \right) ~.
\label{vdef}
\end{eqnarray}
This type of velocity dispersion results from a picture of the vacuum
as a quantum-gravitational `medium', which responds differently to the
propagation of particles of different energies and hence
velocities. This is analogous to propagation through a conventional
medium, such as an electromagnetic plasma~\cite{latorre}.  The
gravitational `medium' is generally believed to contain microscopic
quantum fluctuations, which may occur on scale sizes of order the
Planck length $L_{\P} \sim 10^{-33}$~cm on time scales of order
$t_{\P} \sim 1/ E_{\P}$, where $E_{\P}\sim10^{19}$~GeV.  These
may~\cite{emn,garay} be analogous to the thermal fluctuations in a
plasma, that occur on time scales of order $t \sim 1 / T$, where $T$
is the temperature.  Since it is a much `harder' phenomenon associated
with new physics at an energy scale far beyond typical photon
energies, any analogous quantum-gravity effect could be distinguished
from by its different energy dependence: the quantum-gravity effect
would {\rm increase} with energy, whereas conventional medium effects
{\rm decrease} with energy in the range of interest~\cite{latorre}.

\medskip \par
Equation (\ref{vdef}) encodes a minute modification for most practical
purposes, since $E_{\QG}$ is believed to be a very high scale,
presumably of order the Planck scale $E_{\P}\sim10^{19}$~GeV. Even so,
such a deformation could be rather significant for even
moderate-energy signals, if they travel over very long
distances. According to (\ref{vdef}), a signal of energy $E$ that
travels a distance $L$ acquires a `time delay', measured with respect
to the ordinary case of an energy-independent speed $c$ for massless
particles:
\begin{eqnarray}
 \Delta t \sim \xi {E \over E_{\QG}} {L \over c}  ~.
\label{delayt}
\end{eqnarray}
This is most likely to be observable when $E$ and $L$ are large whilst
the interval $\delta\,t$ over which the signal exhibits time structure
is small. These are the respects in which GRBs offer particularly good
prospects for such measurements, as we discuss later.

\medskip \par 
Before doing so, we first review briefly how modified laws for the
propagation of particles have emerged independently in different
quantum-gravity approaches.  The suggestion that quantum-gravitational
fluctuations might modify particle propagation in an observable way
can already be found in~\cite{ehns,emn}. A phenomenological
parametrization of the way this could affect the neutral kaon
system~\cite{ehns,emnk,hp} has been already tested in laboratory
experiments, which have set lower limits on parameters analogous to
the $E_{\QG}$ introduced above at levels comparable to
$E_{\P}$~\cite{CPLEAR}. In the case of massless particles such as the
photon, which interests us here, the first example of a
quantum-gravitational medium effect with which we are familiar
occurred in a string formulation of an expanding
Robertson-Walker-Friedman cosmology~\cite{aben3}, in which photon
propagation appears tachyonic. Deformed dispersion relations that are
consistent with the specific formula (\ref{vdef}) arose in approaches
based on dimensionful ``$\kappa$'' quantum deformations of Poincar\'e
symmetries~\cite{kpoin}. Within this general class of deformations,
one finds~\cite{kpoin,gacxt} an effect consistent with (\ref{vdef}) if
the deformation is rotationally invariant: the dispersion relation for
massless particles $c^2{\bf p}^2 = E_{\QG}^2 \, \left[ 1 -
e^{E/E_{\QG}}\right]^2$, and therefore $\xi = 1$.  
It should be noted that a deformed dispersion relation has
also been found in studies of the quantization of point particles in a
discrete space time~\cite{thooft}.

\medskip \par
A specific and general dynamical framework for the emergence of the
velocity law (\ref{vdef}) has emerged~\cite{aemn1} within the
Liouville string approach~\cite{emn} to quantum gravity, according to
which the vacuum is viewed as a non-trivial medium containing
`foamy' quantum-gravity fluctuations.  The reader can visualize the
nature of this foamy vacuum by imagining processes that include the
pair creation of virtual black holes.
Within this approach, it is possible to verify that massless particles
of different energies excite vacuum fluctuations 
differently as they propagate through 
the quantum-gravity medium, giving rise to
a non-trivial dispersion relation of
Lorentz `non-covariant' form, just as in a thermal medium. 
The form
of the dispersion relation is not known exactly, but its structure
has been studied~\cite{aemn1} via a perturbative expansion,
and it was shown in~\cite{aemn1} that 
the leading $1/E_{\QG}$ correction is in agreement
with (\ref{vdef}).

\medskip \par
It has been recently suggested~\cite{garay} the vacuum might have
analogous `thermal' properties in a large class of quantum-gravity
approaches, namely all approaches in which a minimum length $L_{\rm min}$, 
such as the Planck length $L_{\P} \sim 10^{-33}$ cm,
characterizes short-distance physics. These should in general lead to
deformed photon dipersion relations with $E_{\QG} \sim 1/L_{\rm min}$, 
though the specific form (\ref{vdef}) may not hold in all
models, and hence may be used to discriminate between them. In support
of (\ref{vdef}), though, we recall~\cite{aemn1,gacxt} that this type
of non-trivial dispersion in the quantum-gravity vacuum has
implications for the measurability of distances in quantum gravity
that fit well with the intuition emerging from recent heuristic
analyses~\cite{gacmpla} based on a combination of arguments from
ordinary quantum mechanics and general relativity.

\medskip \par
We now explain how GRBs provide an excellent laboratory for testing
such ideas, now that the cosmological origin of at least some of them
has been established.  We recall that typical photon energies in GRB
emissions are in the range $0.1-100$~MeV~\cite{review}, and it is
possible that the spectrum might in fact extend up to TeV
energies~\cite{TeV}. Moreover, time structure down to the millisecond
scale has been observed in the light curves~\cite{review}, as is
predicted in the most popular theoretical models~\cite{threv}
involving merging neutron stars or black holes, where the last stages
occur on the time scales associated with grazing orbits. Similar time
scales could also occur in models that identify GRBs with other
cataclysmic stellar events such as failed supernovae Ib, young
ultra-magnetized pulsars or the sudden deaths of massive
stars~\cite{threv2}. We see from equations (\ref{vdef}) and
(\ref{delayt}) that a signal with millisecond time structure in
photons of energy around 20~MeV coming from a distance of order
$10^{10}$ light years, which is well within the range of GRB
observations and models, would be sensitive to $E_{\QG}$ of order
$10^{19}$~GeV $\sim 1/ L_{\P}$.

\medskip \par
Significant sensitivities may already be attainable with the present
GRB data. Sub-millisecond time-structure has been seen in
GRB~910711~\cite{newmicroburst}, and a recent time-series
analysis~\cite{microburst} of the light curve of GRB~920229 using the
Bayesian block technique has identified a narrow microburst with a
rise and decay timescale of order $100\,\mu$sec.  This is seen {\em
simultaneously} in three of the four energy channels of the BATSE
detector on the Compton Gamma Ray Observatory, covering the energy
regions 20-50~keV, 50-100~keV and 100-300~keV respectively. From the
time structure of this microburst we think it should be possible to
extract an upper limit of $\Delta\,t \lappeq 10^{-2}$~sec on the
difference in the arrival times of the burst at energies separated by
$\Delta\,E\sim\,200$~keV. If a burst such as this were to be
demonstrated in future to lie at a redshift $z\sim\,1$, as seems quite
plausible, the implied sensitivity would be to $E_{\QG} \sim
10^{16}$~GeV, and it would be possible to improve this to $\sim
10^{18}$~GeV if the time difference could be brought down to the rise
time reported in~\cite{microburst}. We note in passing that the
simultaneous arrival of photons of different energies from such a
large distance also imposes an upper limit of order $10^{-6}$~eV on a
possible photon mass, but this is much less stringent than other
astrophysical and laboratory limits \cite{pdg}.

\medskip \par
These levels of sensitivity are even more interesting in light of the
fact that recent theoretical work on quantum gravity, particularly
within string theory, appears to favor values of the effective scale
characterizing the onset of significant quantum-gravity effects that
are somewhat below the Planck scale, typically in the range $10^{16}$
to $10^{18}$~GeV~\cite{Mtheory}. If our scale $E_{\QG}$ were indeed to
be given by such a novel effective quantum-gravity scale, parts of the
GRB spectrum with energies around $0.1$~MeV and millisecond time
structure (or energies of order 100~MeV and 1-second time structure,
or energies around 1~TeV and 1-hour time structure) might be sensitive
to the type of candidate quantum-gravity phenomenon discussed in this
paper.

\medskip \par 
In order to provide some quantitative comparison of the GRB
sensitivity to this phenomenon with that of other astrophysical
phenomena, we can compare values of the ``sensitivity factor'' $\eta
\equiv {|\Delta t^*|/\delta\,t}$, where $\delta\,t$ represents the
time structure of the signal, whilst $\Delta\,t^*$ is the time delay
acquired by the signal if $E_{\QG}\sim\,E_{\P}$, namely $\Delta
t^*\sim\pm{E\,L/(c\,E_{\P})}$. As already discussed, GRB emission with
millisecond time structure and energy around $20$ MeV that travels a
distance of order $10^{10}$ light years has $\eta\sim1$. Another
interesting possibility is that we may observe lensing of a GRB by a
foreground galaxy~\cite{lens,nemiroff}.  The burst would then reach us
by two or more different paths whose light travel times would differ
typically by weeks up to years. Since conventional gravitational
lensing is achromatic, any {\em energy-dependence} in the time delay
would be a direct probe of the new physics of interest, and would be
independent of the actual emission mechanism of gamma-ray bursts.  We
note that the HEGRA~\cite{hegra} and Whipple~\cite{whipple} air
\u{C}erenkov telescopes have already searched for TeV emission from
the direction of GRBs, motivated by the EGRET detection~\cite{egret}
of emission up to 18~GeV from GRB~940217. If such searches were to
prove successful and moreover identified a lensed GRB, one would be
able to infer via (\ref{delayt}) a sensitivity to $\eta \sim 10^{-6}$.

\medskip \par
As observed in~\cite{aemn1}, which did not consider GRBs whose
cosmological distances were not then established, pulsars and
supernovae are among the other astrophysical phenomena that might at
first sight appear well suited for probing the physics we are
interested in here, in light of the short time structures they
display. However, although pulsar signals have very well defined time
structure, they are at relatively low energies and are observable over
distances of at most $10^4$ light years. If one takes an energy of
order $1$ eV and postulates generously a sensitivity to time delays as
small as $1~\mu$sec, one estimates a sensitivity to $\eta
\sim10^{-10}$. With new experiments such as AXAF it may be possible
to detect X-ray pulsars out to $10^6$ light years, allowing us to
probe up to $\eta \sim 10^{-8}$.

\medskip \par
Concerning supernovae, we observe that neutrinos from Type II events
similar to SN1987a, which should have energies up to about 100~MeV
with a time structure that could extend down to milliseconds, are
likely to be detectable at distances of up to about $10^{5}$ light
years, providing sensitivity to $\eta\sim10^{-4}$. We have also
considered the cosmic microwave background. Although the distance
travelled by these photons is the largest available, the only possible
signature is a small distortion of the Planck spectrum due to the
frequency-dependence of $c$, which we estimate to be of order
$\Delta\,I(\nu)/I(\nu)\sim\,\nu/E_{\P}\sim10^{-32}$, which is quite
negligible.

\medskip \par
We conclude that, in principle, GRBs allow us to gain many orders of
magnitude in the sensitivity factor $\eta$. Moreover, and most
importantly, this high sensitivity should be sufficient to probe
values of the effective scale characterizing the onset of
quantum-gravity effects extending all the way up to the Planck scale,
as illustrated by the estimates we have provided. Ideally, one would
like to understand well the short-time structure of GRB signals in
terms of conventional physics, so that the novel phenomena discussed
here may be disentangled unambiguously. However, even in the absence
of a complete theoretical understanding, sensitive tests can be
performed as indicated above, through the serendipitous detection of
short-scale time structure~\cite{microburst} at different energies in
GRBs which are established to be at cosmological distances. 
Detailed features of burst time series should enable the
emission times in different energy ranges to be put into
correspondence.
Since any time shift due to quantum-gravity effects of the
type discussed here would {\em increase} with the photon energy, this
characteristic dependence should be separable from more conventional
in-medium physics effects, which {\em decrease} with energy. To
distinguish any quantum-gravity shift from effects due to the source,
we recall that the medium effect would be {\em linear} in
the photon energy, which would not in general be the case for time
shifts at the source. To disentangle any such effects, it is clear
that the most desirable features in
an observational programme would be fine time resolution at high
photon energies.

\baselineskip 12pt plus .5pt minus .5pt

\vfill
\noindent {\Large \bf Acknowledgements}

One of us (G.A.-C.) thanks James Binney and other members of the
Oxford Theoretical Astrophysics group for informative discussions.
This work was supported in part by a grant from the Foundation
Blanceflor Boncompagni-Ludovisi (G.A.-C.), a P.P.A.R.C. advanced
fellowship (N.E.M.)  and D.O.E. Grant DE-FG03-95-ER-40917 (D.V.N.).

\end{document}